\documentclass[preprint2]{aastex}

\usepackage{epsf}
\usepackage{graphics}

\newcommand{\chandra}{{\it Chandra }}
\newcommand{\etal}{{\it et al. }}
\shorttitle{High-redshift AGN luminosity function}
\shortauthors{Shankar \& Mathur}

\begin{document}


\def \charthoffset {\hspace{0.2cm}} \def \charthsep {\hspace{0.3cm}}
\def \chartvsepcap {\vspace{0.3cm}}
\def \chartvsep {\vspace{0.1cm}}
\newcommand{\putchartb}[1]{\clipfig{/home/halley/dgrupe/ps/#1}{85}{20}{0}{275}{192}}
\newcommand{\putchartc}[1]{\clipfig{/home/halley/dgrupe/ps/#1}{55}{33}{19}{275}{195}}
\newcommand{\chartlineb}[2]{\parbox[t]{18cm}{\noindent\charthoffset\putchartb{#1}\charthsep\putchartb{#2}\chartvsep}}

\newcommand{\chartlinec}[2]{\parbox[t]{18cm}{\noindent\charthoffset\putchartc{#1}\charthsep\putchartc{#2}\chartvsep}}


\title{On the faint end of the high redshift AGN luminosity function}


\author{Francesco Shankar \& Smita Mathur}
\affil{Astronomy Department, Ohio State University,
    140 W. 18th Ave., Columbus, OH-43210, U.S.A.}
\email{shankar, smita@astronomy.ohio-state.edu}




\begin{abstract}
Using the results of recent optical surveys we conclude that the
{\it non}-detection of quasars down to faint magnitudes implies a
significant flattening of the high redshift ($z\sim 6$) optical
active galactic nuclei (AGN) luminosity function for
$M_{1450}\gtrsim-24.7$. We find that all the data are consistent
with a faint-end slope for the optical AGN luminosity function of
$\beta=-2.2$ and $\beta=-2.8$, at the 90\% and 99\% confidence level
respectively, flatter than the bright-end slope of $\beta'\sim
-3.2$. We also show that X-ray deep surveys have probed even fainter
magnitudes than the optical ones yielding more significant
constraints on the shallow faint-end slope of the optical luminosity
function. The inclusion of Type II AGN candidates, detected in the
\chandra deep fields, hints towards an higher normalization for the
total AGN luminosity function, if these sources lie at $5\lesssim z
\lesssim 6.5$. We then discuss simple theoretical models of AGN
formation and evolution in the context of cold dark matter
cosmology. The comparison with the total AGN luminosity function
favors a redshift-dependent relation between black hole and dark
matter halo masses of the type $M_{\bullet}\propto M_{\rm
halo}^{\alpha}$, with $1.3\lesssim \alpha\lesssim 1.7$, compatible
with independent studies from statistical analysis and
rotation curve measurements. 
Finally we compute the quasar contribution to reionization to be
$\lesssim 9\%$ at $z\sim 6$, up to $\sim 30\%$ when integrated
within $5.5\lesssim z \lesssim 6.5$, significantly smaller than that
from galaxies.
\end{abstract}

\keywords{galaxies: active - quasars:general}

\section{Introduction}
\label{sec|Intro}

A fundamental tool to study the evolution of Active Galactic Nuclei
(AGNs) is their luminosity function (LF) and its evolution with
redshift.  This enables us to probe the accretion history of
Supermassive Black Holes (SMBHs) which, in turn, set constraints on
galaxy evolution, as SMBHs are believed to be ubiquitous and tightly
linked with all stellar bulges and spheroids today (e.g. Ferrarese
\& Ford 2005 and references therein). Moreover constraining the AGN
LF is essential to understand the AGN contribution to the total
emissivity background in different bands, their environmental
impact, their clustering properties and the role AGNs had, if any,
in the cosmological re-ionization of Hydrogen and Helium.

X-ray observations carried out with \chandra, {\it ASCA}, {\it
HEAO-1} are statistically complete in X-ray to a level of $\sim
90\%$ up to $z\sim 3--4$ (e.g. La Franca \etal 2005). These
observations have assessed that AGNs statistically follow a
luminosity-dependent density evolution (e.g. Ueda et al. 2003), in
which lower luminous AGNs peak at $z\lesssim 1$, while higher
luminous ones, usually optically bright, peak at higher redshifts,
$z\sim 2$ (e.g. Osmer 2004). However how far in time can we track
AGN evolution? The picture is still not yet complete at higher
redshifts, where X-ray measurements of the AGN LF are lacking.

On the optical side the SDSS team (Richard \etal 2006), and earlier
works by Kennefick \etal (1995), COMBO-17 (Wolf \etal 2003) have
probed the bright tail of the optical AGN LF up to $z\sim 4.7$ down
to magnitudes of $M_{1450}\lesssim -23.3$ showing that a simple
power-law can easily fit the data. The GOODs team (Cristiani \etal
2004; Fontanot \etal 2006a) has been able to set constraints on the
shape of the optical LF down to $M_{1450}\sim -21.5$ and up to
$z\sim 5$, showing evidence for a flattening at the faint-end. Fan
\etal (2004; F04 from here on) have instead probed the very luminous
tail of the $z\sim 6$ optical LF up to $M_{1450}\lesssim -26.5$.

Here we use existing observations of optically selected AGN and
\chandra deep fields to determine constraints on the redshift 6
shape of both the optical and X-ray LF. We will in particular show
that X-ray observations allow us to probe the $z\sim 6$ LF to $\sim
2$ magnitudes fainter than the deepest optical observations. In
section~\ref{sec|Method} we describe our method, present the
constraints on the faint end LF in section~\ref{sec|results}, their
implications in sections~\ref{sec|implic} and
~\ref{sec|reionization}, and conclude in section~\ref{Concl}. We use
concordance cosmological parameters $\Omega_{m}=0.27$,
$\Omega_{\Lambda}=0.73$, $\Omega_{\rm b}=0.022h^{-2}$, $H_{0}=71$ km
s$^{-1}$ Mpc$^{-1}$, $\sigma_8=0.84$, and $\Gamma=0.2$ (Spergel
\etal 2003).

\section{Method}
\label{sec|Method}

In the following we will always refer to the total population of
active galactic nuclei as AGNs. We will call "QSOs" (or "quasars")
the type I, optical AGNs, and Type II AGNs those which have a column
density above $\log N_H/\rm{cm}^{-2}=22$. The latter kind of sources
are often very faint or even missed in optical surveys, but still
detected in X-rays.

Several deep X-ray and optical surveys have now probed several
portions of the AGN luminosity function at high redshifts, however
often finding a scarce number of faint sources. In this paper we
will show that the paucity of sources at these faint fluxes gives
the most up-to-date stringent constraints on the knee and the
faint-end slope of the QSO LF and give hints to the overall AGN
distribution at $z\sim 6$. Similar studies, though with looser upper
limits, have been carried out by Sharp \etal (2004), Willott \etal
(2005), Wyithe (2004) and Dijkstra \& Wyithe (2006).

To convert X-ray to optical rest-frame luminosities we will use the
equation (see Richards \etal 2005)
\begin{equation}
\log L_{\rm 2\, keV}=\log L_{\rm 2500}+\alpha_{\rm ox}\log
\left(\frac{\nu_{\rm 2\, keV}}{\nu_{\rm 2500}}\right)\, ,
    \label{eq|aoxLxLo}
\end{equation}
between monochromatic luminosities and a constant photon index
$\Gamma=1.8$ to compute the broad-band X-ray luminosities. Many
authors (e.g. Vignali \etal 2003) have found that the spectral index
$\alpha_{\rm ox}$ depends significantly on luminosity. In
particular, the latest results by Steffen \etal (2006), who have
collected a large number of optically and X-ray selected AGNs at
several redshifts, claim a significant correlation of the spectral
index with luminosity and a marginal one (1.1$\sigma$) with reshift,
i.e.
\begin{eqnarray}
\alpha_{\rm ox}=(-0.126\pm 0.013)\log L_{\rm 2500}-(0.01\pm \\
0.009)z+(2.311\pm 0.372)\, , \nonumber
    \label{eq|aoxVSLAndz}
\end{eqnarray}
where $L_{2500}$ is the monochromatic luminosity at $2500\, {\AA}$,
in units of ${\rm erg\, s^{-1}\, Hz^{-1}}$.

In the following we will use results from several optical/NIR
surveys carried out on different areas. To compare the luminosities
probed by different observations, we will first need to convert the
broadband z$_{\rm AB}$, where most of the observations are carried
out, to the apparent magnitude at wavelength $1450(1+z){\AA}$, i.e.
AB$_{1450(1+z)}$, and finally convert the latter to the rest-frame
luminosities M$_{1450}$.

We convert z$_{\rm AB}$ to AB$_{1450(1+z)}$ magnitude following the
spectrum described in Cristiani \& Vio (1990) and updated in
Cristiani \etal (2004). To convert then to absolute magnitudes
M$_{1450}$, we note that in the AB system $m_{\rm AB}= -2.5\log
f_{\nu}-48.6$, where the monochromatic flux $f_{\nu}$ is in units of
${\rm erg\, cm^{-2}\, s^{-1}\, Hz^{-1}}$, and the absolute magnitude
is M$_{\rm AB}=-2.5\log L_{\nu(1+z)} +51.6$, where L$_{\nu(1+z)}$ is
the luminosity at rest frame frequency $\nu(1+z)$, in units of ${\rm
erg\, s^{-1}\, Hz^{-1}}$. Using $f_{\nu}= (1+z)L_{\nu(1+z)} /4\pi
D^2_{\rm L}(z)$ leads to (see also Stern \etal 2000b)
\begin{eqnarray}
  {\rm M}_{1450}=AB_{1450(1+z)} + 2.5\log (1+z) + 97.45\\
  - 5\log D_{\rm L}(z) \nonumber \label{eq|MzMbM1450}
\end{eqnarray}
and (Fan \etal 2001)
\begin{equation}
{\rm M}_{\rm B}={\rm M}_{1450} + 2.5 \alpha \log(4400/1450) + 0.12
\label{eq|MbM1450}
\end{equation}
%
where ${\rm M}_{\rm B}$ is the absolute B magnitude calibrated for a
$f_{\nu}\propto \nu^{\alpha}$ (where we set $\alpha=-0.5$) spectrum
and the factor 0.12 comes from the zero point difference between the
AB and Vega-based system (Fan \etal 2001). 

To constrain the shape of the $z\sim 6$ AGN LF, following Willott
\etal (2005), we first define a double power-law model LF of the
type
\begin{eqnarray}
 \Phi(M_{1450},z)=\frac{dn(M_{1450},z)}{dM}=\rho(z)\times \\
\frac{\Phi_0}{10^{0.4(\beta+1)(M_{1450}-M_{1450}^*)}+10^{0.4(\beta'+1)(M_{1450}-M_{1450}^*)}},
\nonumber
    \label{eq|modelLF}
\end{eqnarray}
where $\rho(z)=10^{-0.48(z-z_0)}$, with $z_0=6.07$, is the average
redshift of the F04 LF. With such a parametrization we assume that
the high-redshift AGN LF evolves with a Pure Density Evolution (PDE)
beyond redshift of 5 at the same rate as the LF in Fan \etal (2001).
Following the observations by F04, $\Phi_0$ has been calibrated to
match the cumulative number density of $\sim 6.4\times 10^{-10}\,
{\rm Mpc^3}$ for sources with $M_{1450}<-26.7$ at $z=z_0$ (these are
the limit and number density for $\Omega_{m}=0.35$,
$\Omega_{\Lambda}=0.65$, $H_{0}=65$ and have been converted to our
cosmology). The bright-end slope and the break magnitude have been
fixed to $\beta'=-3$ and $M_{1450}^*=-24.67$ respectively, as this
provides a very good match to the Cool et al. 2006 observations (see
below). The faint-end slope $\beta$ has been left variable and
constrained by matching the observed number density of sources in
each survey. At $z\sim 5$ the LF in equation~(\ref{eq|modelLF}) is
in agreement with the cumulative number density determined by Fan
\etal (2001).

Thus given a model LF we integrate it within the effective volume of
the survey using the following equation (see e.g. Fan \etal 2001)
\begin{eqnarray}
n_{\rm obs}=\Delta \Omega\int_{z_{\rm min}}^{z_{\rm max}} dz
\frac{dV}{dzd\Omega}\times\\
\int_{L_{\rm min}(z)}^{\infty} \Phi(L,z)dL \, \rm{p}(L,z,SED)\, ,
\nonumber
    \label{eq|integralLF}
\end{eqnarray}
and tune the faint-end slope $\beta$ in equation~(\ref{eq|modelLF})
to match the observed number of sources $n_{\rm obs}$ in the survey.
$L_{\rm min}(z)$ refers to the redshift dependent absolute
luminosity, corresponding to the minimum optical apparent magnitude
(or the minimum flux) considered. We will in general assume a
constant selection function p(L,z,SED) for each survey, as indicated
by the authors or derived from their papers. In the following we
will always use the form in equation~(\ref{eq|modelLF}) for
$\Phi(L,z)$, unless in some specified cases, in which we will adopt
the Fan et al. (2001) LF. The latter is a single power-law function
valid in the range $4<z<5.5$, in which we let the slope $\beta_0$ to
be variable, while we keep the redshift evolution and normalization
at $M_{1450}^*$ as in Fan \etal (2001; both $M_{1450}^*$ and
normalization have been converted to our cosmology, which is
required at each redshift bin during the evolution).

\section{Results}
\label{sec|results}

\subsection{The Type I AGN Luminosity Function}
\label{sec|TypeILF}

A significant improvement with respect to previous works has been to
fix the knee of the LF in equation~(\ref{eq|modelLF}) to the value
recently inferred by Cool \etal (2006). These authors discovered
three objects down to z$_{\rm AB}$=21.93, (i.e $M_{1450}\sim -24.67$
at z$=5.53$ using equation~(\ref{eq|MzMbM1450})) in the AGN and
Galaxy Evolution Survey (AGES) spectroscopic observations of the
NOAO Deep Wide-Field Survey (NDWFS). They covered 7.7 deg$^2$ of the
NDWFS field with an average completeness level of 54\% down to
$I_{\rm AB}=22$ mag. These authors claim their observations agree
with the F04 LF with a slope of $\beta=-3.2$ and a redshift
evolution as in Fan et al. (2001) and in
equation~(\ref{eq|modelLF}). Using equation ~(\ref{eq|integralLF}),
we find $\sim 3$ sources with z$_{\rm AB}\gtrsim 21.93$ at $z>5$, in
very good agreement with observations and with the authors. The
integration was pursued up to $z=7$, as beyond this limit the
optical selection of quasars becomes impossible because the
Ly$\alpha$ line redshifs out of the z$_{\rm AB}$-band (see Cool
\etal 2006).

Following a similar procedure we find significant evidence for a
flattening of the QSO LF below $M_{1450}\sim -24.67$. Recently,
Mahabal \etal (2005, M05 hereafter) in the field around the $z=6.42$
quasar SDSS J1148+5251 discovered a quasar at redshift $z=5.70$ with
luminosity of $M_{1450}\sim -23.6$, i.e. z$_{\rm AB}=23.0$. This
quasar is the faintest known optically selected quasar at $z>5.5$,
allowing determination of the faint end of the quasar LF to fainter
luminosities.

The authors claim the source, identified in a $\sim 2.5$ deg$^2$
survey, has a surface density consistent with the Wolf \etal (2003)
LF extrapolated through a PDE within their effective volume, which
follows the $R$-band drop-out criterion of Stern \etal (2000), i.e.
$4.3\lesssim z\lesssim 5.8$. We have checked that adopting the
latter determination for the effective volume, we recover the
results of the authors only if an average $\sim 28\%$ selection
detection probability is used. However using the Wolf \etal (2003)
LF, which is defined only for $z \lesssim 4.7$, needs an
extrapolation to higher redshifts. Moreover the functional form in
equation~(\ref{eq|modelLF}) has been calibrated on the data by Cool
\etal (2006) at $z\gtrsim 5$. We therefore replace
equation~(\ref{eq|modelLF}) with the Fan \etal (2001) QSO LF, more
appropriate for the redshift bin probed by the $R$-band drop-out
technique.

\begin{figure*}[t!]
\resizebox{\hsize}{!}{\includegraphics[clip=true]{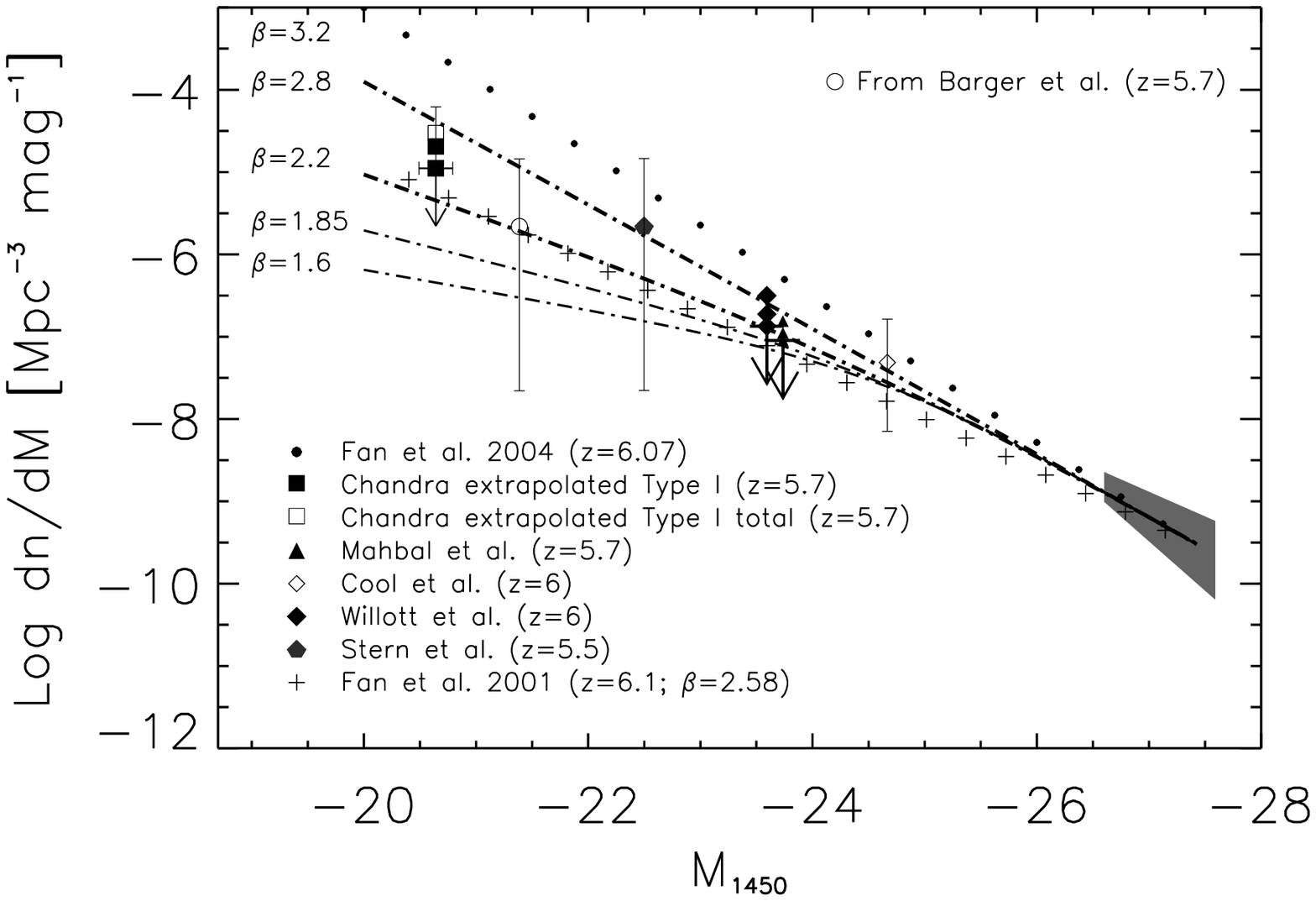}}
\caption{\footnotesize Type I AGN Luminosity Function at $z\sim 6$.
The {\it squares} represent the estimated AGN density at the optical
magnitude of $M_{1450}\sim -20.6$ needed for a single source to be
detected above the broad band X-ray flux limit of $f_{\rm 0.5-8\,
keV}\approx 2.5\times 10^{-16}\, {\rm erg\, cm^{-2}\, s^{-1}}$ in
160 arcmin$^2$. From top to bottom the two {\it solid squares} are
the 99\% and 90\% confidence level estimates, while the {\it open
square} has been estimated taking into account all the nine
spectroscopically unidentified sources in the deep \chandra fields
with its 99\% error bars; the horizontal error bar reflects the
uncertainty in the computation of the optical magnitude at fixed
X-ray flux when the 1-$\sigma$ uncertainty in
equation~(\ref{eq|aoxVSLAndz}) are taken into account. The {\it open
circle} is the AGN luminosity function estimated from Barger \etal
(2003) with its 99\% CL. The {\it solid pentagon} is the AGN density
from Stern \etal (2000) with the 99\% confidence level error bars;
the {\it solid diamonds} are the AGN densities from Willott \etal
(2005) again shown with 99\%, 95\% and 90\% confidence level (from
top to bottom); the {\it solid triangles} are the AGN densities
computed from Mahabal \etal (2005), with 99\%, 95\% and 90\%
confidence level (from top to bottom); the {\it open diamond} is the
AGN density computed from Cool \etal (2006) with its 99\% confidence
level error bars; the {\it crosses} are the Fan \etal (2001)
luminosity function extrapolated to $z=6.1$ assuming PDE from $z=5$.
The values reported on the upper-left of the figure are the absolute
values for the faint slopes of the model AGN luminosity function in
equation~(\ref{eq|modelLF}); the \emph{shaded} area is the observed
luminosity function by Fan \etal (2004) with its $1-\sigma$
uncertainty while the {\it thick-dotted} line is its extrapolation
to fainter magnitudes.} \label{fig|LFz6}
\end{figure*}

Keeping the same value for the selection function we find that the
detection of one object in the effective volume probed by M05, is
consistent with a number density at $M_{1450}\sim -23.6$ of
$\log[dn/dM]\sim -6.70$ at the 99\% confidence level (CL; with
$\beta_0=-2.8$), down to $\log[dn/dM]\sim -6.95$ at the 90\% CL
(with $\beta_0=-2.54$; see Table~\ref{Table|Survey}; we plot the M05
data in Figure~\ref{fig|LFz6} shifted by -0.15 magnitudes for
avoiding overcrowding with Willott et al.'s data). Their data, shown
as solid triangles in Figure~(\ref{fig|LFz6}), are consistent with
extrapolating equation~(\ref{eq|modelLF}) to fainter optical
magnitudes using $\beta=-2.3$ and $\beta\sim -1.7$ at the 99\% and
90\% CL respectively (see Table~\ref{Table|Survey}). Note that these
are conservative estimates. Extending the computation of the LF down
to the limiting magnitude of the M05 survey (z$_{\rm AB}=24.5$,
$M_{1450}\sim -22$)
or using a higher completeness function would yield even lower limits on $\beta$.  

Willott \etal (2005) undertook a multicolor quasar optical imaging
with the Canada-France-Hawaii Telescope. They found a lack of
quasars down to a limiting magnitude of z$_{\rm AB}=$23.35 (i.e.
$M_{1450}\sim -23.63$ at $z=6.1$; see Figure~\ref{fig|LFz6}). The
few detected quasar candidates had been subsequently rejected as
being identified with low-mass stars in the near-infrared follow-up
imaging. Using equations~(\ref{eq|modelLF}) and
~(\ref{eq|integralLF}), we find that the Willott \etal (2005)
results are consistent with a faint-end slope of $\beta=-2.95$ at
99\% CL, down to $\beta=-2.27$ at the 90\% CL. We use the selection
function computed from simulated colors for high-redshift QSOs in
the redshift range $5.5\lesssim z\lesssim 6.7$ (see Figure 2 in
Willott \etal 2005). This is the most conservative selection
function from Willott \etal (2005), determined for very extreme red
color cuts. Using a less conservative estimate (solid line in their
Figure 2) our results drop to $\beta=-2.84$ at 99\% CL, down to
$\beta=-2.2$ at the 90\% CL.

The above findings are shown in Figure~\ref{fig|LFz6}. The gray area
shows the data with its $1 \sigma$-error bars by F04 for the $z\sim
6.1$ LF, while the solid dots show the extrapolation of the F04 LF
to faint luminosities assuming a single slope of $\beta=-3.2$. The
presence of a significant downturn in the QSO LF for magnitudes
below $M_{1450}\sim -24.67$ is evident. In particular the M05
observations are significantly stringent in constraining the
faint-end slope.

\begin{deluxetable}{ccccccc}
\tablecolumns{6} \tablewidth{0pc} \tablecaption{Faint-end slopes for
the QSO LF at several confidence levels} \tablehead{{survey} &
\colhead{$|\beta|$} & \colhead{90\%} & \colhead{95\%} &
\colhead{99\%} & \colhead{Area} & \colhead{p}}\startdata
Mahabal et al. &  & 1.7    &  1.9 & 2.3   & 2.5 sq. deg.     & 28 \%\\
Willott et al. &  & 2.27    &  2.57 & 2.95  & 3.32 sq. deg.    & variable\\
Barger et al. (single source) &  & 2.32   &  2.38 & 2.48  & 160 arcmin$^2$   & 50 \%\\
\enddata
\tablecomments{All the values in the first three columns give the
faint-end slopes $\beta$ (equation~(\ref{eq|modelLF})) needed to
reproduce the observed number density in each survey; all confidence
levels are computed through Poissonian statistics following Gehrels
\etal (1986); the column before last indicates the effective area
probed by each survey; the last column indicates the average
selection function used for each survey.} \label{Table|Survey}
\end{deluxetable}

In Figure~\ref{fig|LFz6} we also show the interesting result by
Stern \etal (2000). These authors have discovered a quasar at
$z=5.50$ from deep, multicolor, ground-based observations covering
74 arcmin$^2$ with z$_{\rm AB}=23.4$. To estimate their average
selection function we redo the calculations in Stern \etal (2000)
and integrate down to $M_B=-22.5$ (in their cosmology) the Boyle
\etal (1991) $z=2$ quasar LF, evolved as in Schmidt \etal (1995).
The effective volume probed is again defined by the $R$-band
drop-out as in M05 (Stern \etal 2000). We find that an average
completeness of p$=0.89$ recovers their quoted $\sim$0.15 quasars in
the field of view. As for M05, we then use the more up-to-date Fan
\etal (2001) LF, more suitable for the redshift range probed by this
survey, and equation~(\ref{eq|integralLF}). Even in this case we let
the slope to be variable, $\beta_0=-2.92$ in the specific case,
while we keep the redshift evolution and normalization at
$M_{1450}^*$ as in Fan \etal (2001).

We integrate down to $AB_{1450(1+z)}=24.1$ (see Table 1 in Stern
\etal 2000), instead of using the measured z$_{\rm AB}=23.4$. This
is because the authors find their source to have unusual spectral
features for which they derive a special continuum model which is
significantly different from the average spectral template used in
this paper. The detection of one source in their field of view is
consistent with a comoving density of $\log [dn/dM(-22.5,5.5)]\sim
-5.66\, {\rm Mpc^{-3}\, mag^{-1}}$, lower than the F04 extrapolation
at the $\gtrsim$ 80\% CL (see Figure~\ref{fig|LFz6}). The number
density we find still shows evidence for a turnover in the QSO LF
expressed in equation~(\ref{eq|modelLF}), with $\beta=-2.8$, though
less significant than the above results, in agreement with the
authors.

We now turn to the analysis of X-ray data. Barger \etal (2003; see
also Cowie \etal 2003) have used deep multicolor optical data to
search for $z>5$ AGNs in the 2 Ms X-ray \chandra Deep Field-North
exposures. They found 500 sources in their field of view out of
which 423 have z$_{\rm AB}<25.2$. Color analysis was carried out on
all the sources, while 249 sources have spectroscopically confirmed
redshifts. Of these, only one object was found at $5<z<6$ and none
at $z>6$. They found a faint AGN at $z=5.19$ with z$_{\rm AB}=23.9$,
corresponding to an absolute magnitude of $M_{\rm 1450}\sim -22.8$,
i.e. nearly a magnitude fainter than the M05 source, showing the
power of X-ray observations. Barger \etal (2003) claim that none of
the spectroscopically unidentified sources with z$_{\rm AB}<25.2$ is
red enough to have $z>5$. With the aid of the $V/V_{\rm max}$ method
the authors estimated the LF in the comoving volume $5<z<6.5$ to be
consistent with the extrapolation of the Fan \etal (2001) LF with a
slope of $\beta=-2.6$ in the luminosity bin $43 \le \log L_{2-10\,
{\rm keV}}\le 44$ (in ${\rm erg\, s^{-1}}$).
%
%

Here we proceed with a similar calculation but using
equation~(\ref{eq|integralLF}) and extrapolating the Fan \etal
(2001) LF at redshifts beyond $z\sim 5$, in the same manner we
computed the number density for Stern \etal (2000). We find that the
number of
observed sources down to z$_{\rm AB}=25.2$ 
integrated between $5\lesssim z \lesssim 6.5$, with $\beta=-2.1$ and
a completeness of p$=0.7$ (Cowie \etal 2002) is $\sim 1$, in very
good agreement with observations. We considered an area of 160
arcmin$^2$, corresponding to the solid angle where the hard X-ray
flux limit has a signal-to-noise ratio of S/N=3 (see Figure 19 in
Alexander \etal 2003). The X-ray flux considered here is $f_{\rm
0.5-8\, keV}\approx 2.5 \times 10^{-16}\, {\rm erg\, cm^{-2}\,
s^{-1}}$ (Alexander \etal 2003; Barger \etal 2003). This flux limit
is the minimum level of X-ray flux which corresponds to an optical
magnitude brighter than z$_{\rm AB}=26.9$ (the optical detection
limit in Barger \etal (2003)) within the whole effective volume
considered. This is required as all the 500 sources in the CDFN have
optical colors.

We find a number density of $\log$ $[dn/dM$ $(-21.4,5.7)]$ $\sim
-5.70 \, ({\rm Mpc^{-3}\, mag^{-1}})$ corresponding to $\log
L_{2-10\, {\rm keV}}$ $\sim 43.6 \, ({\rm erg\, s^{-1}})$
(equation~(\ref{eq|aoxVSLAndz})), a factor of $2--3$ higher, but
broadly consistent, with the results by Barger \etal (2003; plotted
in Figure~\ref{fig|LFz6} as an open circle with its 99\% CL error
bars).

Indeed deep X-ray observations effectively allow us to probe even
fainter optical luminosities. Mathur \etal (2002; Brandt \etal 2002)
detected three $z\sim 6$ quasars discovered by Fan \etal (2001b)
with \chandra with exposure times of less than 10 ksec. The objects
are detected in X-rays up to 55 keV in rest frame, demonstrating the
incredible sensitivity of \chandra to detect faint sources. The
faintest quasar in their sample, SDSS 1030$+$0524, at $z=6.28$ was
detected in about 8 ksec with an observed flux of $f_{\rm 0.5-8\,
keV}=4\times 10^{-15}\, {\rm erg\, cm^{-2}\, s^{-1}}$
. The \chandra deep field observations have exposure time of 2 Msec,
so could detect down to $2000/8=250$ times fainter sources with a
flux limit of $f_{\rm 0.5-8\, keV}\approx 1.6\times 10^{-17}\, {\rm
erg\, cm^{-2}\, s^{-1}}$. This however would be the sensitivity
limit at the aim point, close to the aim point sensitivity reported
by Alexander \etal (2003). Therefore we expect the \chandra
observations to be sensitive in detecting sources down the minimum
flux considered, i.e., we can extend the computation using
equations~(\ref{eq|modelLF}) and ~(\ref{eq|integralLF}) down to the
limit of $f_{\rm 0.5-8\, keV}\approx 2.5\times 10^{-16}\, {\rm erg\,
cm^{-2}\, s^{-1}}$ (i.e. down to z$_{\rm AB}\sim 26$ at the average redshift of $z=5.7$).
%
In Figure~\ref{fig|LFz6} and Table~\ref{Table|Survey} we report our
findings. Using a completeness of about 50\%, as computed by Cowie
\etal (2002) for faint fluxes (notice that the use of an higher
completeness level would strengthen our results), we find that the
detection of a single source in the \chandra Deep Field in the range
$5<z<6.5$, is consistent with a faint slope of $\beta=-2.48$ at the
99\% CL, down to $\beta=-2.32$ at the 90\% CL (shown as solid
squares).

Barger \etal (2003) claim that none of the unidentified sources is
red enough to be at $z>6$. They however also point out that up to
nine out of the optically faint spectroscopically unidentified
sources with z$_{\rm AB}>25.2$, could still possibly lie in the
range $5<z<6.5$. Therefore assuming these sources to rely in the
same luminosity bin as the single $z=5.19$ source, they find an
upper limit to their comoving LF about an order of magnitude higher
than their previous estimate. Following the same approach undertaken
for the case of a single source, we apply
equations~(\ref{eq|modelLF}) and ~(\ref{eq|integralLF}) down to
$f_{\rm 0.5-8\, keV}\approx 2.5\times 10^{-16}\, {\rm erg\,
cm^{-2}\, s^{-1}}$ and using a completeness of $\rm{p}=0.5$, proper
for faint fluxes (Cowie \etal 2002). We consider all the nine
sources not to be heavily obscured, i.e. they are Type I AGN. We
find that the comoving number density of the LF to be $\log
[dn/dM(-20.6,5.7)]\sim -4.52 \, ({\rm Mpc^{-3}\, mag^{-1}})$, using
a slope $\beta=-2.57$ (shown as an open square with its 99\% CL in
Figure~\ref{fig|LFz6}). Our estimate is higher but still consistent
with the one by Barger \etal (2003). To notice also that our
estimate is also quite conservative, lower limits would be inferred
extending the computation to lower magnitudes, close to the survey
limit (z$_{\rm AB}=26.9$).

The 1-$\sigma$ uncertainties in equation~(\ref{eq|aoxLxLo}) produce
uncertainties of $\sim 0.15$ magn in the final computation of the
optical $M_{\rm 1450}$ magnitude, as shown in Figure~\ref{fig|LFz6}.
We also notice that our conclusions are quite insensitive to the
actual fit used for the optical-to-X-ray flux ratios. For example,
similar results are also obtained using the relation between
luminosity in the rest-frame $R$ band, $L_R$, and the hard X-ray
luminosity $L_{2-10\, {\rm keV}}$ calibrated by La Franca \etal
(2005; see their equation (4)) for Type I AGNs. The Vignali \etal
(2003) fit instead of equation~(\ref{eq|aoxLxLo}) would yield even
fainter optical magnitudes for the same X-ray flux and number
density, thus strengthening our results for a presence of
a flattening at the faint-end of the QSO LF. 

We conclude this section noting that all the data are consistent
with a break in the QSO LF at $M_{1450}\gtrsim -24.67$. The
faint-end slope of the {\it optical} Type I AGN LF, is consistent
with being on average $\beta \sim -2.2$, at the 90\% CL and $\beta
\sim -2.8$, at the 99\%
CL (see Figure~\ref{fig|LFz6}). 
Our overall findings are consistent with the Fan \etal (2001) LF
with $\beta=-2.58$ extrapolated to $z=6.1$ (shown with "plus" signs
in Figure~\ref{fig|LFz6}). This clearly shows how the high-redshift
LF is more consistent with a PDE beyond $z\sim 5$. Richards \etal
(2006) have recently found from a sample of SDSS QSOs, that the
high-redshift QSO LF bright slope flattens at $z\gtrsim 2.5$ from
$|\beta'|=3.2$ down to $|\beta'|\gtrsim 2.4$ at $z\sim 5$. We
suggest instead that a double power-law, steeper at bright
magnitudes and shallower at fainter ones, is more indicative for the
QSO LF at $z\sim 6$. Our results on the optical LF are instead in
good agreement with those by Fontanot \etal (2006a), who compiled
the QSO LF from a sample of GOODs and SDSS QSOs in the range
$3.5<z<5.2$. They find significant evidence for a bright-end slope
of $|\beta'|=3.3$, a break in the LF at $M_{1450}^*\sim -25$ and a
strong flattening below this limit.

\subsection{The Type I plus Type II AGN Luminosity Function}
\label{sec|TypeIILF}

In the previous section we have considered the detected sources in
the CDFN as Type I AGNs. The single high-redshift source detected
with spectroscopic redshift of $z=5.19$ can in fact be safely
considered as a Type I AGN. Converting its measured hard X-ray flux
$f_{\rm 2-8\, keV}\sim 7\times 10^{-16}\, {\rm erg\, cm^{-2}\,
s^{-1}}$ to optical magnitudes (using
equation~(\ref{eq|aoxVSLAndz})) we get z$_{\rm AB}\sim 23.6$, very
close to its actual magnitude of z$_{\rm AB}\sim 23.9$. However some
of the nine optically faint spectroscopically unidentified objects
found by Barger \etal (2003) could be considered moderately obscured
sources, with very faint optical fluxes (though still detectable).
In this section we give constraints on the faint-end shape for the
AGN LF, including the CDFN spectroscopically unidentified sources
assuming that a certain fraction of them are Type II AGNs lying at
$z\sim 6$.

Indeed the $2-8\, {\rm keV}$ band corresponds to the $14-56\, {\rm
keV}$ intrinsic band at $z=6$, which is not much affected by
absorption for column densities $\log N_H \lesssim 24\, ({\rm
cm^{-2}})$. Here we notice however that we do not expect the
detection of any very obscured, Compton-thick source in the CDFN,
with $\log L_{2-10\,{\rm keV}}\lesssim 44\, ({\rm erg\, s^{-1}})$.
From Wilman \& Fabian (1999) we have taken the spectrum of two
typical AGNs with column densities $\log N_H=24.25\, ({\rm
cm^{-2}})$ and $\log N_H=24.75\, ({\rm cm^{-2}})$, derived from a
Monte-Carlo realization which includes Compton down-scattering which
reduces the intensity even in the highest bins of energy. We then
computed the ratio between absorbed and unabsorbed luminosities (the
former is the actual \emph{observed} flux) for the various spectra
in the $z=6$ redshifted, restframe energy hard band $14-70\, {\rm
keV}$ (the least absorbed band of the CDFN).

We find that the ratio $L_{14-70\, {\rm keV}}^{\rm abs}/L_{14-70\,
{\rm keV}}^{\rm unabs}$ is $\sim 71\%$ for $\log N_H=24.25\, ({\rm
cm^{-2}})$ and $\sim 30\%$ for $\log N_H=24.75\, ({\rm cm^{-2}})$
(to notice for comparison that an AGN with $\log N_H=23.5\, ({\rm
cm^{-2}})$ has a ratio of $\sim 96\%$). A typical source with
intrinsic luminosity $L_{2-10 \,\rm keV}=10^{44}\, ({\rm erg\,
s^{-1}})$ (fainter luminosities would be below the X-ray flux limit)
in the observed frame will in turn produce observed fluxes $f_{\rm
2-8\, keV}\approx 2\times 10^{-16}\, {\rm erg\, cm^{-2}\, s^{-1}}$
and $f_{\rm 2-8\, keV}\approx 1\times 10^{-16}\, {\rm erg\,
cm^{-2}\, s^{-1}}$ for respectively the $\log N_H=24.25\, ({\rm
cm^{-2}})$ and $\log N_H=24.75\, ({\rm cm^{-2}})$ case. Therefore
sources with $\log N_H$ $\gtrsim 24\, ({\rm cm^{-2}})$ and $\log
L_{2-10\,{\rm keV}}\lesssim 44\, ({\rm erg\, s^{-1}})$, would be
hardly visible in the CDFN field of view (we remind that the flux
limit is $f_{\rm 2-8\, keV}\approx 2\times 10^{-16}\, {\rm
erg\, cm^{-2}\, s^{-1}}$).

On the other hand moderately obscured sources (with $\log
N_H\lesssim 24\, ({\rm cm^{-2}})$) would definitely be visible at
these high redshifts, and some of them might have been already
discovered in the deep \chandra fields, among the optically faint
Barger \etal (2003) sample. Other possible indications of the
presence of Type II sources in the high redshift universe also come
from Koekemoer \etal (2004). From the 2 msec {\rm Hubble} Deep
Field-North and the 1 msec CDFS, these authors have robustly
detected a sample of seven sources with extreme optical/X-ray flux
ratios $>$10 at a flux limit above $f_{\rm 0.5-8\, keV}\approx
3\times 10^{-16}\, {\rm erg\, cm^{-2}\, s^{-1}}$. Although these
kind of sources are generally rare, if they are AGNs lying at $z>6$
such that their Ly$\alpha$ emission has been redshifted out of the
z$_{\rm AB}$ bandpass, then they would definitely be high-redshift
obscured sources. We cannot directly make a prediction with their
data as we are converting to optical luminosities and these are
special sources, strong outliers with respect to
equation~(\ref{eq|aoxVSLAndz}) (also a conversion to bolometric
luminosity would be difficult in this case).

In this section we still adopt the same X-ray $K$-correction as for
unabsorbed sources with $\Gamma=1.8$ to convert flux to
luminosities. We still use equation~(\ref{eq|aoxVSLAndz}), which
holds for Type I AGNs, for converting X-ray fluxes to intrinsic
optical luminosities for Type II AGNs, assuming that unification
models are on average valid. First we consider \emph{all} the
sources detected by Barger et al. (2003) as optically faint Type II
AGNs lying at $z>5$. We then integrate equation~(\ref{eq|modelLF})
in the redshift range $5\lesssim z \lesssim 6.5$ above a flux limit
of $f_{\rm 2-8\, keV}=2\times 10^{-16}\, {\rm erg\, cm^{-2}\,
s^{-1}}$ corresponding to an effective area of 40 arcmin$^2$ (see
Figure 19, bottom panel in Alexander et al. 2003). The adopted flux
limit corresponds to apparent magnitudes z$_{\rm AB}>25.5$ at $z
\sim 5.7$ (equation~(\ref{eq|aoxVSLAndz})). Using $\beta=3.45$ we
obtain $\sim 9$ sources in the field of view, corresponding to a
comoving density of $\log [dn/dM(-21.2,5.7)]\sim -3.85 \, {\rm
Mpc^{-3}\, mag^{-1}}$.

\begin{figure*}[t!]
\resizebox{\hsize}{!}{\includegraphics[clip=true]{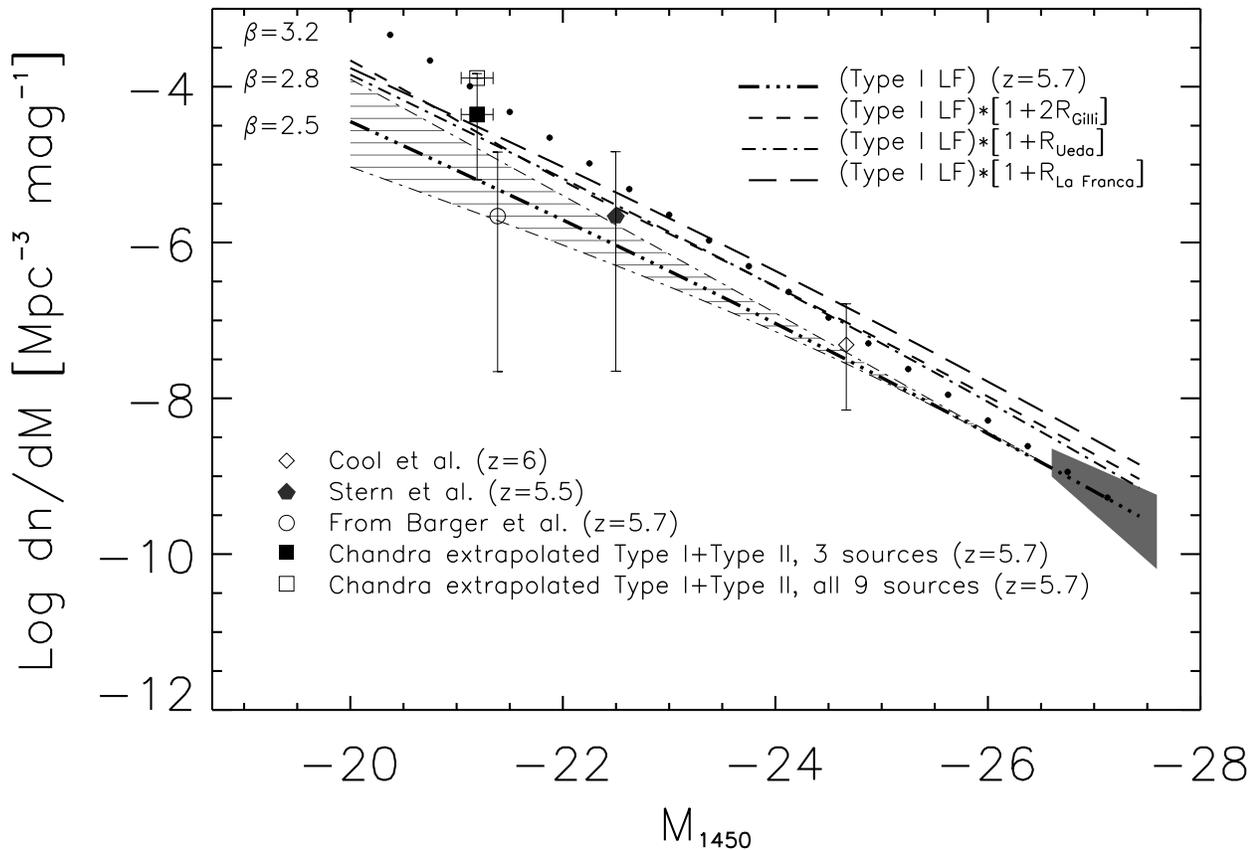}}
\caption{\footnotesize Type I plus Type II AGN luminosity function.
The {\it open square} is the AGN luminosity function considering all
the nine sources in the \chandra deep fields as Type II AGNs; the
{\it solid square} is instead the AGN luminosity function computed
when 3 sources are taken from the Barger\etal (2003) optically faint
sample to be $z\sim 6$ AGN candidates. The {\it triple dot-dashed}
line is the Type I AGN luminosity function with the average slope of
$\beta=-2.5$ and the \emph{striped} area is the range of 90--99\% CL
for the Type I AGN luminosity function; the {\it dashed} line is the
Type I AGN luminosity function multiplied by $1+2 R$ where $R$ is
the Type II/Type I ratio as given in Gilli \etal (2006;
equation~(\ref{eq|Ratio})), while in the \emph{dot-dashed} and
\emph{long-dashed} lines $R$ has been extracted from Ueda \etal
(2003) and La Franca \etal (2005) respectively. All the other
symbols and lines are as in Figure~\ref{fig|LFz6}.}
\label{fig|LFz6obscu}
\end{figure*}

The results are shown with an open square with its 99\% confidence
level in Figure~\ref{fig|LFz6obscu} with its error bar on the
optical luminosity when using equation~(\ref{eq|aoxVSLAndz}). If the
nine sources are confirmed to be Type II AGNs at $z\sim 6$, this
would show evidence for an extraordinary contribution of Type II
AGNs at the faint end of the AGN LF. Assuming all these faint
sources to lie at $z\sim 6$ would imply a fraction
obscured-to-unobscured AGNs of order 9. X-ray surveys at lower
redshifts found much more moderate numbers for this ratio (e.g. Ueda
\etal 2003, La Franca \etal 2005, Gilli \etal 2006).

In fact the above scenario is an extreme one, as most probably many
of these optically faint sources probably belong to lower redshifts
(see discussions in Barger \etal 2003 and Cowie \etal 2003). 
Therefore we more safely assume that up to 3 sources, out of the 9
detected, i.e. three times the surface density of the Type I LF
(made up of only the single $z=5.19$ source), are actual
contributors to $z\sim 6$ AGN LF with $\log N_H\lesssim 24\, {\rm
cm^{-2}}$, following the results by La Franca et al. (2005), who
find an almost flat distribution of the fraction of sources per
logarithm bin of $N_H$. In this second case we get a comoving
density of $\log [dn/dM(-21.2,5.7)]\sim -4.3 \, {\rm Mpc^{-3}\,
mag^{-1}}$, shown as a solid square in Figure~\ref{fig|LFz6obscu}
with its 99\% CL error bars.

We then compare our AGN LF estimate with what would be obtained by
multiplying the average QSO LF (with a faint slope of $\beta=-2.5$;
section~\ref{sec|TypeILF}) by a factor $1+R$, where $R$ is the ratio
between Type II and Type I AGNs, as calibrated by several groups for
AGNs detected at lower redshifts. For computing $R$ we also include
sources up to $\log N_H\lesssim 26\, {\rm cm^{-2}}$. We use the
results by Ueda \etal (2003), La Franca \etal (2005) and Gilli \etal
(2006). As specified above, we consider Type II AGNs all those
sources with $\log N_H> 22\, {\rm cm^{-2}}$. Therefore by definition
the number ratio Type II-to-Type I will be given by $R=n[22<\log
(N_H/{\rm cm^{-2}}) <26]/n[\log (N_H/{\rm cm^{-2}})<22]$. For Ueda
et al. (2003) and La Franca et al. (2005) we use their quoted
distributions for the fraction of objects up to $\log N_H=26\, {\rm
(cm^{-2})}$.

Gilli \etal (2006) developed an updated model for the synthesis of
the X-ray background, for which they adopted a different definition
for the number ratio between obscured and unobscured AGNs, i.e.
$R=n[21<\log (N_H/{\rm cm^{-2}}) <24]/n[\log (N_H/{\rm
cm^{-2}})<21]$. According to their model this ratio is
redshift-independent, luminosity-dependent and, translated to
optical magnitudes, it reads as
\begin{eqnarray}
R(M_{1450})=R_s\times \exp \left(-10^{-0.4(M_{1450}-M_c)}
\right)+\\
R_q\times \left(1-\exp \left(-10^{-0.4(M_{1450}-M_c)} \right)
\right)\, , \nonumber
    \label{eq|Ratio}
\end{eqnarray}
with $R_s=3.6$, $R_q=1$ and $M_c=-20.7$, i.e. $\log L_{0.5-2\, {\rm
keV}}\sim 43.5\, ({\rm erg\, s^{-1}})$, where $R_s$ and $R_q$ are
the $n[21<\log (N_H/{\rm cm^{-2}}) <24]$ $/n[\log (N_H/{\rm
cm^{-2}})<21]$ ratio at the low- and high-luminosity regimes
respectively, and $M_c=-20.7$ is the characteristic luminosity
dividing the two regimes (Gilli \etal 2006). To include also
Compton-thick sources with column densities up to $\log N_H=26\,
{\rm (cm^{-2})}$, we consider double this ratio (see Figure 7 in
Gilli \etal 2006).

The resulting AGN LF from the Gilli et al. (2006) model is shown
with a dashed line in Figure~\ref{fig|LFz6obscu}. In the same figure
we also plot the results of using $R$ as calibrated from Ueda \etal
(2003; dot-dashed line) and La Franca \etal (2005; long-dashed
line). All the models for the distribution of sources as a function
of $N_H$ considered provide similar estimates for the total AGN LF,
and are consistent with the extrapolated data at the faint end, when
using 3 sources as proxy for the $z\sim 6$ AGN population within the
effective area. We remind here that our estimate for the faint-end
AGN LF could easily be underestimated, as the CDFN cannot detect
Compton-thick sources (those with $\log N_H>24\, {\rm cm^{-2}}$).

To notice that all the models also provide a significant number of
sources at the bright end of the AGN LF in excess of the QSO LF. Due
to the small volume probed by the CDFN we however cannot provide any
kind of constraint on the actual contribution of intrinsically very
luminous and more rare obscured sources. Anyway in the following we
will consider as our best estimate of the total AGN LF at $z\sim 6$,
the QSO LF multiplied by (1+$R_{\rm Ueda}$), where $R_{\rm Ueda}$ is
the Type II-to-Type I ratio as extracted from Ueda et al. (2003). In
the following section we will compare theoretical models with such a
determination of the AGN LF.

\section{Implications}
\label{sec|implic}

The high redshift AGN LF is a very sensitive tool to probe the
connections between supermassive black holes (SMBHs) and their host
dark matter (DM) halos. In this section we discuss possible physical
implications for SMBH evolution from our estimated AGN LF.
We compute the \emph{total} AGN LF from the formation rates of DM
halos simply as (e.g., Mahmood \etal 2005)
\begin{equation}
\Phi(L_B,z)\simeq t_{\rm AGN}\times R_{\rm MM}(M_{\rm halo},z)\times
\frac{dM_{\rm halo}}{dM_{\bullet}}\frac{dM_{\bullet}}{dL_B}\, .
\label{eq|PhiLz}
\end{equation}
Here $R_{\rm MM}(M_{\rm halo},z)$ is the formation rate of DM halos
with mass $M_{\rm halo}$ at redshift $z$, massive and virialized
enough to host a galaxy/AGN; $t_{\rm AGN}$ in this model is the
average timescale during which an AGN is active (but not necessarily
visible). We neglect any dependence on mass for this timescale and
just set it equal to $t_{\rm AGN}=10^7 {\rm yr}$. $M_{\bullet}$ is
the SMBH mass producing the luminosity $L_B$. In the following for
simplicity we will use extended Press-Schechter (EPS) theory to
compute $R_{\rm MM}(M_{\rm halo},z)$ and we will consider only halos
$\log M_{\rm halo}/M_{\odot}\leq 13.2$. The limiting mass of $M_{\rm
halo}\sim 10^{13} M_{\odot}$ as host for the most massive galaxies
and most luminous quasars has been constrained through clustering
(e.g. Croom \etal 2005), statistical (e.g. Shankar \etal 2006), and
semi-analytical (e.g. Granato \etal 2004) techniques.

Most probably the EPS theory is a rather poor approximation for
describing the statistics of high redshifts AGN host halos, moreover
a more physical treatment of the baryonic processes is required when
comparing with observations (e.g. Lapi \etal 2006, Fontanot \etal
2006b). However here we neglect such issues and focus on the most
suitable relation $M_{\bullet}-M_{\rm halo}$, among those suggested
in the literature, which better compares with observations within
the very basic model expressed in equation~(\ref{eq|PhiLz}). We
therefore consider different recipes for the jacobian relation in
equation~(\ref{eq|PhiLz}) relating DM and SMBH mass.

\begin{figure*}[t!]
\resizebox{\hsize}{!}{\includegraphics[clip=true]{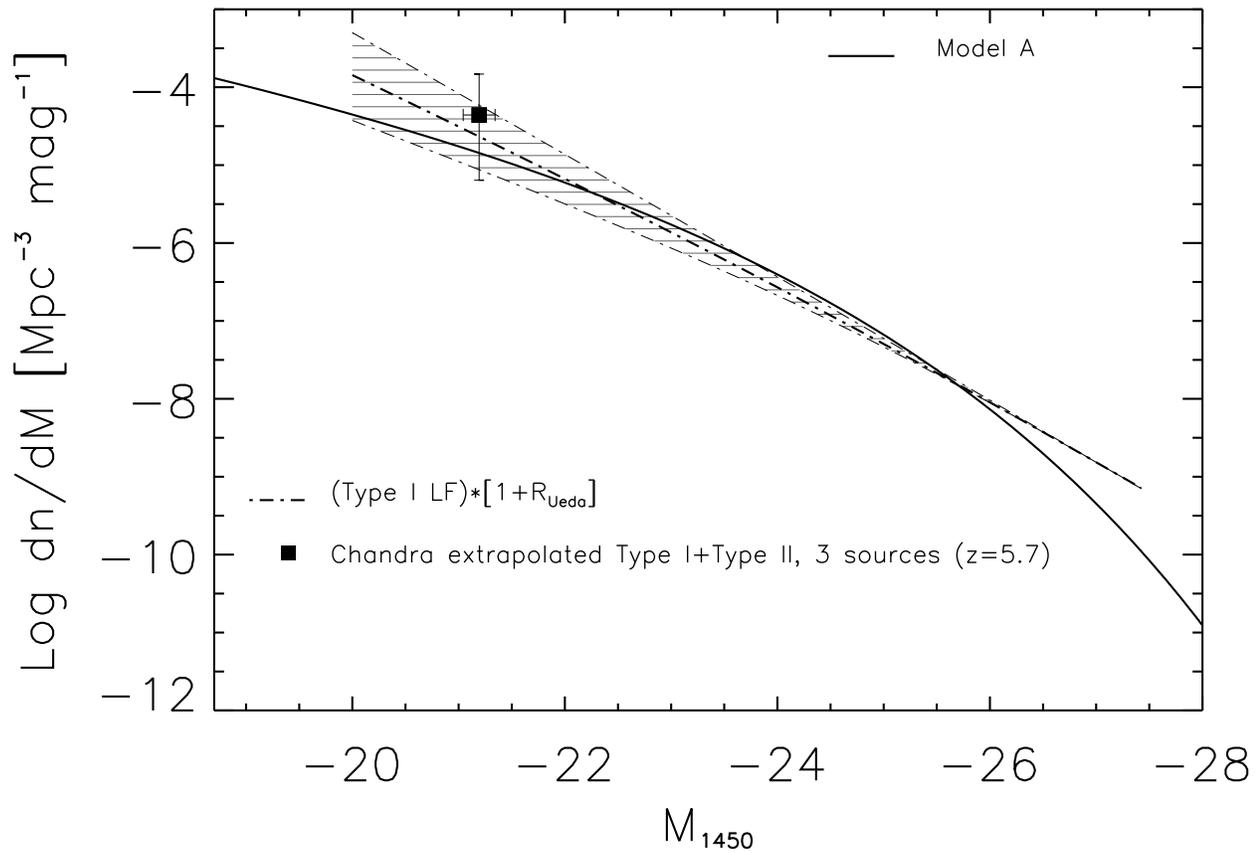}}
\caption{\footnotesize Estimated AGN luminosity function at $z\sim
6$ obtained by considering the average QSO luminosity function
derived in section~\ref{sec|TypeILF} (with its uncertainty region
shown as a \emph{striped} region) and multiplying it by (1+$R_{\rm
Ueda}$), to account for all sources up to $\log N_H=26\, ({\rm
cm^{-2}})$. The {\it solid} line is the result of Model A (see
text); all sources are assumed to radiate at the Eddington limit.}
\label{fig|LFz6Models1}
\end{figure*}

We first discuss the Wyithe \& Loeb (2003; see also Mahmood \etal
2005) model which connects the SMBH and DM halo masses as
$M_{\bullet}\propto M_{\rm halo}^{5/3}(1+z)^{5/2}$, derived from a
basic AGN feedback-constrained model.
We shall refer to this model as model A in the rest of the paper.
Using Model A, we calculate the $z=6.1$ LF, shown as a solid line in
Figure~\ref{fig|LFz6Models1}. This model is consistent with the AGN
LF estimated in section~\ref{sec|TypeIILF}.
Several other groups have also formulated theoretical models for
describing SMBH evolution in the context of cosmological theories
(e.g. Haenelt \& Rees 1993, Cavaliere \& Vittorini 2000, Vittorini
\etal 2005, Hopkins \etal 2005). Here we adopt a simple version of
one of these models, viz. the model by Kauffmann \& Haenelt (2000;
see also Croton 2005) who parameterize the link between SMBH mass
and halo mass as:
\begin{equation}
M_{\bullet}=0.03m_R\left[1+\left(\frac{280\, {\rm km\,
s^{-1}}}{V_{\rm vir}}\right)^2\right]^{-1}\, m_{\rm cold}\, ,
    \label{eq|MbhMhaloKH}
\end{equation}
where $m_R$ represents the merger mass ratio between the central and
satellite galaxies, i.e. $m_R=m_{\rm sat}/m_{\rm central}$, $V_{\rm
vir}$ is the virial velocity of the AGN host halo and $m_{cold}$ is
the amount of cold baryonic matter (which will eventually form
stars). The equation~(\ref{eq|MbhMhaloKH}) prescription with
$m_R=1$, consisting of only major mergers, is our model B.

\begin{figure*}[t!]
\resizebox{\hsize}{!}{\includegraphics[clip=true]{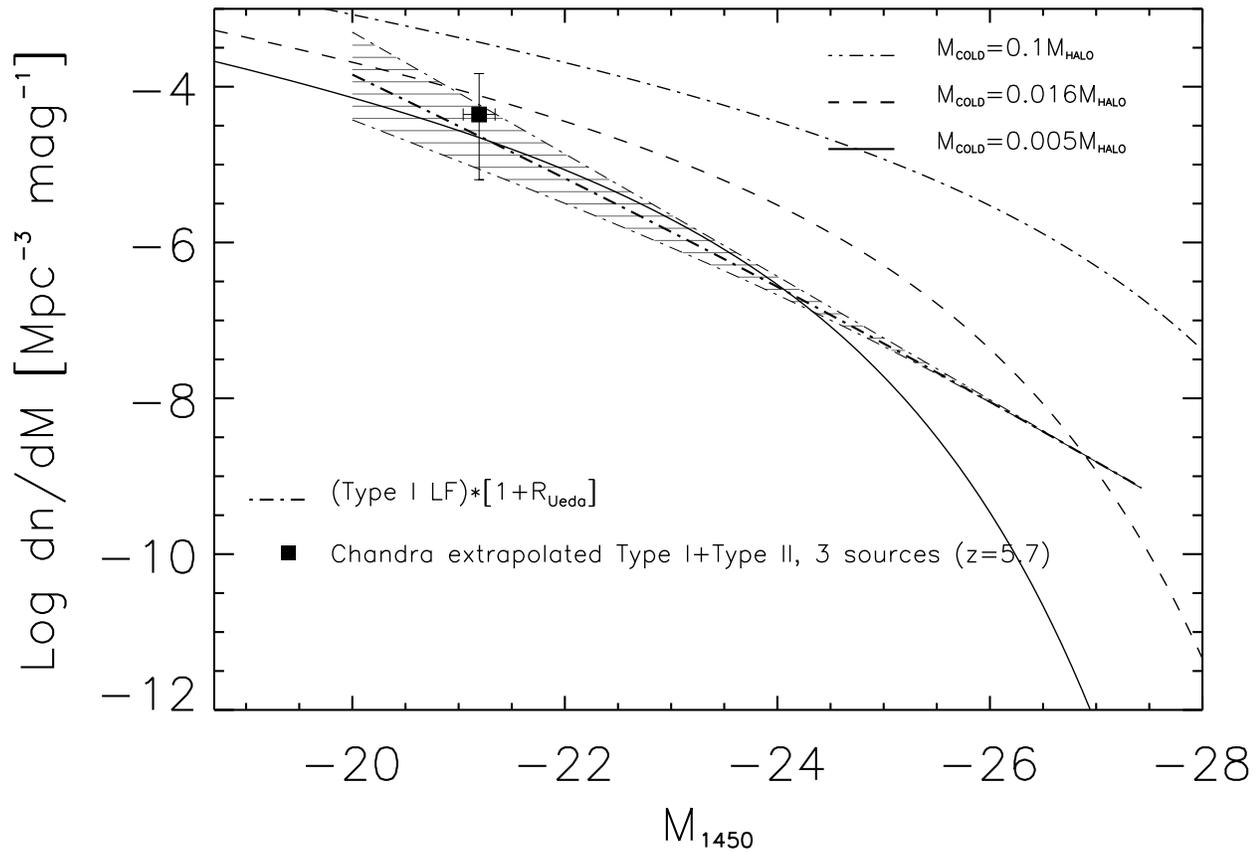}}
\caption{\footnotesize The {\it dot-dashed}, {\it dashed} and {\it
solid} lines are the results of Model B (see text) for different
fraction of cold gas with respect to the dark matter host mass as
labeled; all sources are assumed to radiate at the Eddington limit.
All the other curves are as in Figure~\ref{fig|LFz6Models1}.}
\label{fig|LFz6Models2}
\end{figure*}

We then consider three cases in which the cold fraction is $m_{\rm
cold}= 0.1 \times M_{\rm halo}$, $m_{\rm cold}= 0.016 \times M_{\rm
halo}$, and $m_{\rm cold}= 0.005 \times M_{\rm halo}$. The resulting
LFs are plotted in Figure~\ref{fig|LFz6Models2}. The first case, in
which the amount of baryons per halo is close to the cosmological
value, lies well above the observed LF. The second case fits the
bright-end LF, assuming a baryonic fraction of a few percent,
consistent with results by Shankar \etal (2006). It is interesting
to note that a strong decrease of baryonic mass fraction with
decreasing halo mass is needed to reproduce the AGN LF, down to
only 0.5\% at the very faint end. 

We now test a model based on Ferrarese \etal (2002; see also Shankar
\etal 2006). We call this model C, and is the most empirical way to
track the $M_{\bullet}-M_{\rm halo}$ relation. We relate the SMBH
mass to bulge velocity dispersion using the $M_{\bullet}-\sigma$
relation. The $\sigma$ is then connected to the circular velocity of
the host galaxy by using the empirical $\sigma-V_C$ relation (from
Baes \etal 2003). The circular velocity is then scaled to the virial
velocity assuming a Navarro, Frenk \& White (1996; NFW) profile and
a concentration parameter $c(M_{\rm halo},z)$ from Bullock \etal
(2001). Finally the virial velocity is converted to halo mass using
the virial theorem, $GM_{\rm halo}/R_{\rm vir}=V_{\rm vir}^2(z_{\rm
vir})$ (we use the relations in Loeb \& Peebles 2003 given for each
$z_{\rm vir}$), which is computed at the moment of the virialization
of the host DM halo (i.e. we set $z_{\rm vir}=6.1$). We also add a
scatter of $\sim 0.35$ in the resulting $M_{\bullet}-M_{\rm halo}$
relation, as empirically suggested by Ferrarese (2002) and
theoretically supported by Lapi \etal (2006). Our result reads as
\begin{equation}
M_{\bullet}= 5.8\times 10^7\left(\frac{M_{\rm halo}}{10^{12}\,
M_{\odot}}\right)^{1.39}\left(\frac{1+z}{7}\right)\, M_{\odot}\, ,
    \label{eq|MbhMhaloC}
\end{equation}
which is close to the one calibrated locally through statistical
arguments by Shankar \etal (2006) and to the one obtained by Granato
\etal (2004) and Lapi \etal (2006) in semi-analytical models.

\begin{figure*}[t!]
\resizebox{\hsize}{!}{\includegraphics[clip=true]{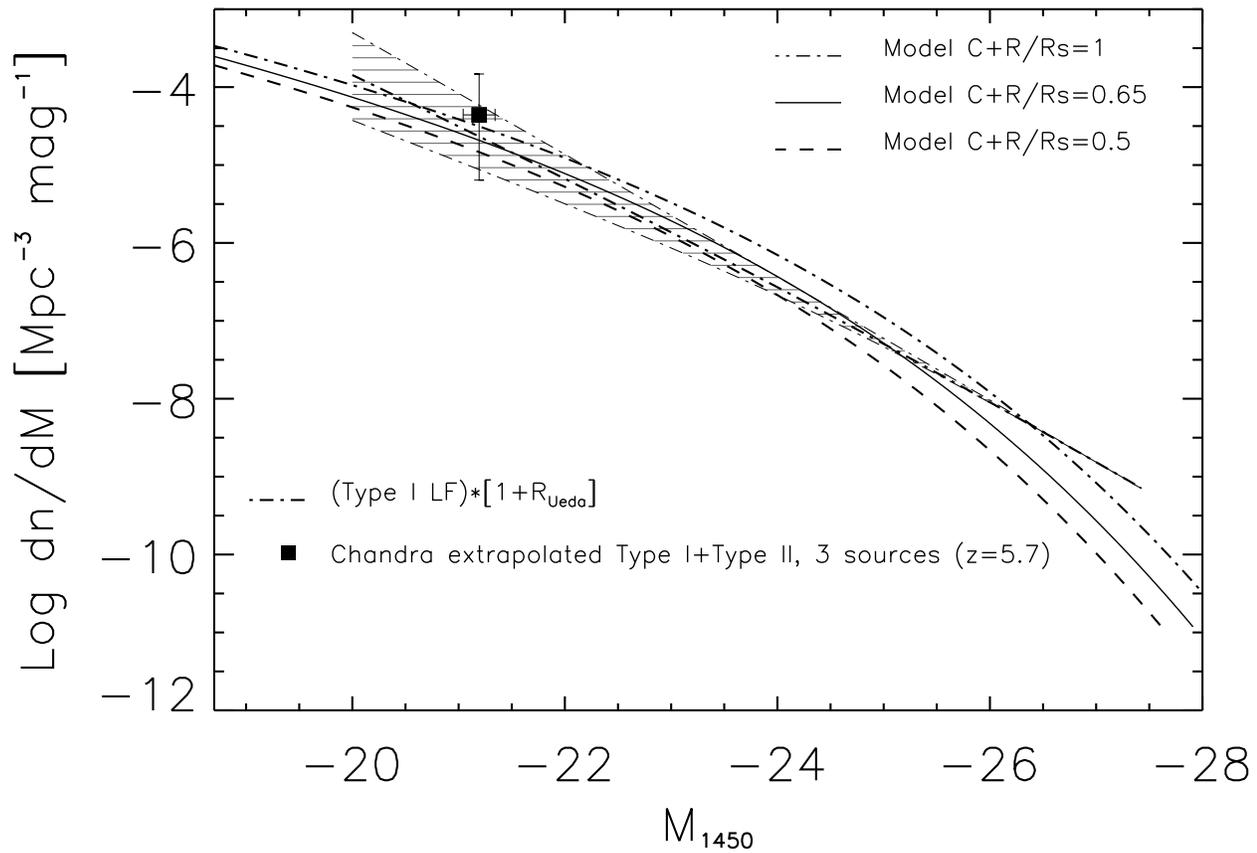}}
\caption{\footnotesize The {\it dot-dashed}, {\it dashed} and {\it
solid} lines are the results of Model C (see text) for different
$R/R_s$ ratios as labeled, assuming that all sources are radiating
at the Eddington limit. All the other curves are as in
Figure~\ref{fig|LFz6Models1}.} \label{fig|LFz6Models3}
\end{figure*}

One uncertainty in the above computations is the radius at which one
measures the circular velocity. Ferrarese (2002) uses $R\sim R_s$
(also used in equation~\ref{eq|MbhMhaloC}), where $R_s$ is the scale
radius of the NFW DM profile.
In Figure~\ref{fig|LFz6Models3} we compare the predictions of model
C with the calibrated high-$z$ AGN LF. The model matches the data
when $R\sim (0.5--1) R_s$ and it has the advantage of being
empirically-based, the only assumption here is that the
$M_{\bullet}-\sigma$ and $\sigma-V_c$ relations are almost constant
in time.

Our findings therefore differ from those in Willott \etal (2005b),
who attempted a similar calculation and found that the number of DM
halos at $z\sim 6$ falls short by several orders of magnitude in
accounting for the number of high-redshift AGNs. Their conclusions
were based on the use of the virial theorem calibrated at $z_{\rm
vir}=0$, which inevitably leads to significantly underestimating the
BH mass (and therefore AGN luminosity) for a given halo. Our
findings are consistent with $\sigma \sim (0.5--0.6)V_{\rm
vir}(M_{\rm halo}, z)$. The latter relation in turn has been used by
Cirasuolo \etal (2005) to show that an impressive fit to the
dispersion velocity function (Sheth et \etal 2003; Shankar \etal
2004) is recovered when integrating the DM formation rates
$R_{MM}(M_{\rm halo}, z_{\rm vir})$ and setting $\sigma \sim 0.57
\times V_{\rm vir}(M_{\rm halo}, z_{\rm vir})$ (see also Loeb \&
Peebles 2003 and Shankar \etal 2006).

The simple approach undertaken in this section shows that a,
redshift-dependent, super-linear relation of the type
$M_{\bullet}\propto M_{\rm halo}^{\alpha}$, with $1.3\lesssim \alpha
\lesssim 1.7$, as the ones proposed in model A and C, is best suited
to represent the AGN LF determined in section~\ref{sec|TypeIILF}.
Alternative scenarios would require a very strong variation of
baryonic content and/or AGN activity timescale with halo mass.

\section{Contribution to Reionization}
\label{sec|reionization}

In this section we compute the possible contribution of AGNs and
stars to the reionization of the universe. In the following we will
consider only Type I AGNs as contributors of the reionization along
our line of sight. The transition from a neutral to a fully ionized
Inter Galactic Medium (IGM) is usually statistically described by a
differential equation for the time evolution of the volume filling
factor of the medium $Q(z)$. The latter quantity quantifies the
level of the IGM {\it porosity} created by the HII, HeII and HeIII
ionization regions around radiative sources such as QSOs and
galaxies.

When the clumping factor of the medium is high, $C>>1$, which is
always a good approximation at high redshifts (the simulations by
Gnedin \& Ostriker 1997 give $C\sim 30$), following Madau \etal
(1999) one can approximate $Q(t)\approx \dot{n}_{\rm ion}/\bar{n}_H
\bar{t}_{\rm rec}$, where $\bar{t}_{\rm rec}$ is the volume-averaged
mean recombination rate, $\dot{n}_{\rm ion}$ is the total rate of
ionizing photons and $\bar{n}_H$ is the mean hydrogen density of the
expanding IGM.

The condition for reionization will then read $Q\approx 1$, which
translates into a total required ionizing photon rate (Madau \etal
1999; Fan \etal 2001):
\begin{equation}
\dot{n}_{\rm
ion}(z)=10^{51.2}\left(\frac{C}{30}\right)\left(\frac{1+z}{6}\right)^3\left(\frac{\Omega_b
h_{0.5}^2}{0.08}\right)^2\, .
    \label{eq|Nion}
\end{equation}

We compare such a rate with those from QSOs and galaxies at similar
redshifts. We start computing the photoionization rate from QSOs,
which is given by (e.g. Madau \etal 1999, Stiavelli \etal 2004)
\begin{equation}
\dot{n}_{\rm AGN}=\int_{\nu_H}^{\infty}\frac{E(\nu)}{h_P\nu}d\nu\, ,
    \label{eq|enuAGN}
\end{equation}
where $h_P$ is the Planck's constant, $\nu_H=3.2\times 10^{15}\,
{\rm Hz}$ is the frequency at the Lyman limit (912 {\AA}, i.e. one
Rydberg). 
The volume emissivity of the total population of sources is given as
\begin{equation}
E(\nu)=\int_{L_{\rm min}}^{\infty}\Phi(L,z)L_{\nu}(L)dL\, ,
    \label{eq|Enu}
\end{equation}
where $L_{\nu}$ is in $\rm erg\, s^{-1}\, Hz^{-1}$. The integration
in equation~(\ref{eq|enuAGN}) is usually performed up to 4 Rydbergs,
as photons more energetic than that are preferentially absorbed by
He atoms. Anyway we have checked that extending the computation to
$\sim 40$ Rydbergs for example, increases our final results by just
$\sim 12\%$. $L_{\rm min}$ in equation~(\ref{eq|Enu}) is instead the
minimum luminosity probed by the survey, which we fix to
$M_{1450}=-20.6$,
following the results in section~\ref{sec|TypeILF}. 

We thus find that the AGN contribution to reionization at redshift
$z\sim 6$ integrating down to $M_{1450}=-20.6$, varies from $2\%$ to
$\sim 9\%$, using a faint-end slope $\beta$ of -2.2, -2.8
respectively. These results are rather independent of the details
for the AGN spectrum. For the computations above we have used the
spectrum described in Schirber \& Bullock (2003), however using, for
example, the three-power law spectrum adopted by Madau \etal (1999),
reduces the AGN contribution by $\lesssim 13\%$.

We follow Meiksin (2005) for computing the photoionization rate from
galaxies
\begin{equation}
\dot{n}_{\rm Galx}=f_{\rm esc}\frac{dN_{\rm phot}}{dM_{\rm
star}}\rho_{\rm star}\, .
    \label{eq|EnuGalx}
\end{equation}
Here $dN_{\rm phot}/dM_{\rm star}\approx 10^{61}\, {\rm phot\,
s^{-1}}$ is the production rate of ionizing photons per solar mass
of stars formed, $\rho_{\rm star}$ is the Star Formation Rate (SFR)
in units of  ${\rm M_{\odot}\, yr^{-1}\, Mpc^{-3}}$, and $f_{\rm
esc}$ is the average fraction of ionizing photons which can freely
escape the star-forming regions into the IGM.

The photons per solar mass has been taken from Smith \etal (2002;
see also Meiksin 2005) for a Salpeter Initial Mass Function (IMF)
and a metallicity of 20\% the solar. 
The SFR was instead recently derived by Richard \etal (2006) to be
$\sim 0.027 {\rm M_{\odot}\, yr^{-1}\, Mpc^{-3}}$ from a deep survey
of lensing clusters aimed at constraining the abundance of
star-forming galaxies at redshift $z\sim 6--10$. We then take
results from the recent paper by Shapley \etal (2006), who have
constrained the escape fraction in the ionizing Lyman continuum from
a sample of 14 $z\sim 3$ star-forming galaxies, finding an average
$f_{\rm esc}\sim (11--18)\%$, and we use the average value of
$f_{\rm esc}\sim 15\%$. In the following we neglect the diffuse
emission from the IGM itself which however could add an additional
photoionization rate (up to 50\%, e.g. Meiksin \& Madau 1993) over
the direct one from the sources. We find that galaxies can
contribute about $50\%$ at $z=6.1$ to reionization. 

We now take a step forward and integrate the contribution to
reionization from both type of sources within the range of the
\chandra deep fields, i.e. $5.5\lesssim z\lesssim 6.5$, and compare
the inferred number of ionizing photons with the number of available
neutral hydrogen atoms at $z\sim 6$, $n_H=6.2\times 10^{66}$ (see
Stiavelli \etal 2004). We find that the contribution from AGNs to
grow up to $(7--30)\%$ while the one from galaxies being about
$70\%$, keeping the SFR constant in this redshift bin, as suggested
by observations. The PDE evolution of the AGN LF for $z<6$ makes the
contribution from AGNs grow more than the one from galaxies.

The photons produced by galaxies reach an average contribution to
reionization of $\gtrsim 70\%$, implying a minor contribution from
QSOs, a conclusion in very good agreement with previous calculations
(e.g. Meiksin 2005, Fan \etal 2001, Barger \etal 2003). However we
also notice that Bouwens \etal (2005) from the deep NICMOS fields in
the Hubble Ultra Deep Fields (HUDF) have detected 5 objects in 5.7
sq. deg., inferring a SFR about a factor of $2--3$ below the one
found by Richard \etal (2006). This would drop the galaxy
contribution down to $(25--40)\%$. Similar points were also recently
addressed by Mannucci \etal (2006) who claim that the high-redshift
SFR could fall short of completely reionizing the universe. They
however also point out that the Bouwens \etal (2005) estimate should
be further increased by a factor of 2 for dust extinction and/or
accounting for luminosity evolution in the estimate. Using therefore
the Bouwens \etal (2005) result in equation~(\ref{eq|EnuGalx}) would
still leave room for a significant contribution from QSOs. Other
ways to fill the possible "missing" photons from galaxies could be
to alter significantly the photon luminosity from massive stars,
invoking extra IGM diffuse light, or including other photoionizing
sources (such as Population III stars and/or miniquasars powered by
intermediate-mass BHs). However the contribution from the latter
sources must be $\lesssim 30\%$ at the most.

\section{Conclusions}
\label{Concl}

In a recent paper, Cool \etal. (2006) found three QSOs at high
redshifts ($z>5.4$) down to magnitudes of ${\rm M}_{1450}\sim
-24.7$. The detection of these quasars is roughly consistent with
the extrapolation of the bright steep slope of the quasar LF by F04
with $\beta=-3$.
At lower magnitudes not many QSOs are observed, as recently
concluded by the optical/NIR surveys conducted by M05, Stern \etal
(2005) and Willott \etal (2005). In this paper we show that such
{\it non}-detection of QSOs implies a significant downturn in the
optical QSO LF at lower magnitudes with respect to those probed by
Cool \etal (2006). Moreover we show that the \chandra deep field
observations probe about $1--2$ magnitudes deeper than the optical
ones (down to ${\rm M}_{1450}= -20.6$). The inferred faint end
co-moving number density is lower by at least an order of magnitude
with respect to the extrapolation of the F04 LF, at a high
significance level.

If a fraction of the nine \chandra deep field sources are Type II
AGNs detected at X-ray fluxes $f_{\rm 2-8\, keV}\gtrsim 2\times
10^{-16}\, {\rm erg\, cm^{-2}\, s^{-1}}$ and with very faint optical
magnitudes (z$_{\rm AB}>25.2$), this would place a constraint on the
shape of the Type I plus Type II AGNs at faint magnitudes, being
consistent with a slope close to the extrapolation of the F04 LF,
i.e. $|\beta|\gtrsim 3$. This estimate of the AGN LF is however
representative of only the sources with $\log N_H\sim 24\, ({\rm
cm^{-2}})$, as more obscured sources would be below the CDFN flux
limit. X-ray surveys at lower redshifts (e.g. La Franca et al. 2005)
have shown that an almost flat distribution in the number of sources
is observed up to column densities of $\log N_H\sim 24\, ({\rm
cm^{-2}})$. We can therefore safely assume that only 3, out of the
nine optically faint sources to be AGNs at $z\sim 6$. This yields a
total AGN LF consistent with updated synthesis models for the XRBG.
%
%

We then compare this result with predictions from three simple
physically and empirically motivated basic models for AGN evolution
in DM halos.
We generally find that a redshift-dependent relation of the kind
$M_{\bullet}-M_{\rm halo}^{\alpha}$, with $1.3\lesssim \alpha
\lesssim 1.7$, produces models more consistent with the AGN LF. Such
relations were also found from independent statistical studies by
Shankar \etal (2006) and rotation curve extrapolations in spirals
and ellipticals (Ferrarese 2002). Alternative models, such as
varying the amount of cold gas fraction or AGN active timescale in
smaller halos, would require extreme fine tuning and a large
variation in the parameters with dark matter halo mass. Here we also
notice that if we reduce the AGN duty-cycle just to the visible
optical phase, i.e. substituting $t_{\rm AGN}$ with $t_{\rm QSO}$ in
equation~(\ref{eq|modelLF}), the match between models and the Type I
AGN LF derived in section~\ref{sec|TypeILF}, would yield similar
results on the type of $M_{\bullet}-M_{\rm halo}$ relation.
%
%

Finally we have shown that AGNs can account for at most $30\%$ of
the total reionization of the Universe at redshifts around $z \sim
6$. However we also notice that a strong caveat on the actual role
of galaxies in reionizing the universe, is set by the normalization
of the cosmological SFR at high redshifts, which is still uncertain
by factors of about $2$.

\begin{acknowledgements}
We acknowledge C. S. Kochanek, P. S. Osmer, D. Weinberg and P.
Martini for helpful discussions. FS also acknowledges S. Cristiani
and F. Fontanot for sharing data on the quasar template spectrum.
The authors are highly indebted to the referee for his/her very
useful comments and suggestions. This work was supported in part by
NASA grant GRT000001640.
\end{acknowledgements}


%

\end{document}